\documentstyle[pra,aps,twocolumn]{revtex}
\def\half{\frac{1}{2}}
\def\<{\langle}
\def\>{\rangle}
\def\Domega{\Delta\omega}

\begin{document}
\title{The Quantum state diffusion model
and the driven damped nonlinear 
oscillator} 
%===============================================================

\author
{M. Rigo\thanks{Email: m.rigo@rhbnc.ac.uk. } $^{^{\hbox{\tiny(a)}}}$, 
G. Alber $^{^{\hbox{\tiny(b)}}}$, 
F. Mota-Furtado $^{^{\hbox{\tiny(a)}}}$ 
and P.F. O'Mahony $^{^{\hbox{\tiny(a)}}}$ \vspace{3mm} \\
$^{\hbox{\tiny(a)}}${\protect\small\em Department of Mathematics, 
Royal Holloway, University of London }\\
{\protect\small\em Egham, Surrey TW20 OEX, United-Kingdom} \vspace{3mm} \\
$^{\hbox{\tiny(b)}}${\protect\small\em Theoretische Quantendynamik, 
Fakult\"at f\"ur Physik, Universit\"at Freiburg }\\
{\protect\small\em D-79104 Freiburg im Breisgau, Germany}
}

\date{ Accepted for publication Phys. Rev. A}

\maketitle

\begin{abstract}
We consider a driven damped anharmonic oscillator which classically
leads to a bistable steady state and to hysteresis.  The quantum
counterpart for this system has an exact analytical solution in the
steady state which does not display any bistability or hysteresis.  We
use quantum state diffusion theory to describe this system and to
provide a new perspective on the lack of hysteresis in the quantum
regime so as to study in detail the quantum to classical transition.
The analysis is also relevant to measurements of a single periodically
driven electron in a Penning trap where hysteresis has been observed.
\end{abstract}

\begin{center}
{\tt PACS 03.65.Bz, 42.50.Lc, 05.30Ch }
\end{center}

\section{Introduction}
%=====================

Emergence of classical properties through interaction with the
environment has been the subject of extensive studies. In this
context, a study of simple open quantum systems can provide a clue to
understanding the mechanism of the quantum-classical transition. In
this paper we wish to study one of the simplest nonlinear quantum
systems - the anharmonic oscillator. This system, when it is damped
and driven, exhibits different behaviour in the steady state when
described either quantum mechanically or classically with the later
showing bistability ~\cite{LL,Kaplan82}.  In order to describe the
emergence of classicality, such differences have to be clarified and
the two results reconciled.

The anharmonic oscillator is of particular interest for several
reasons.  First, its simplicity allows any complexity related to the
model to be avoided. Second, it is the archetypical model for dealing
with nonlinearities in quantum mechanics and has been widely used to
describe a great variety of systems. In particular, it was introduced
by Drummond and Walls~\cite{DW80} to describe dispersive optical
bistability and more recently has been used by Bortman and
Ron~\cite{BortmanRon94,BortmanRon95} to study the relativistic motion
of a resonantly driven electron in a Penning trap.

Drummond and Walls~\cite{DW80} derived the exact steady state
expectations values of the photon distribution function hence showing
that in the quantum regime this model does not exhibit bistability or
hysteresis. 
%change 1 begin
%Hence, the discrepancy between 
%quantum and classical descriptions is manifest. 
%change 1 end
One would expect the classical or quantum
results to be recovered in the appropriate limit.  For instance, the
level of excitation, $\<a^\dagger a\>$, has been used to define such a
limit~\cite{BortmanRon95}. For low excitation number the quantum
result should apply while for high excitation number the classical
result is expected to give the correct behavior.  This description is
reasonable but it does not demonstrate how the transition from
classical to quantum occurs when the excitation number takes an
intermediate value and whether the system suddenly becomes bistable or
whether the bistability appears smoothly.

Drummond and Walls~\cite{DW80} state that the extent to which
bistability is observed will depend on the quantum fluctuations, which
in turn determine the time for random switching between the two stable
states. In this case the classical-quantum transition is of a
statistical nature and the classical result appears in the limit where
the switching time can be considered very large compared to an
observational time.

Bortman and Ron~\cite{BortmanRon94,BortmanRon95} presented a
quantum-mecha\-nical description of a {\em single atomic system} which
is accessible to experiment~\cite{TsengGabrielse95}. They also state
that if one does not deal with the effect of fluctuations upon the two
stable states, that is, its influence upon the time scale of the
stability, bistability is not destroyed by quantum fluctuations and
should be observed even in the case of a low level of
excitation. According to this description the system is bistable in
both regimes and the observation of bistability thus depends on the
experimental set up, with no fundamental restriction on its
observation.

%change1a begin
Motivated by these controversial points of view of
Drummond and Walls~\cite{DW80}
and Bortman and Ron~\cite{BortmanRon94,BortmanRon95}
in this paper
the quantum-classical
transition of the driven damped  anharmonic oscillator 
is investigated
with the help of the quantum state diffusion (QSD) method. 
%change1a end
%change 2 begin
In the context of typical quantum optical problems, for example,
the QSD method
describes the continuous monitoring of the state of a photon source
by individual photoelectric detection processes which involve
heterodyning with a classical intense photon source \cite{C93,WM93}.
However, the QSD method has also been proposed as a phenomenological 
theoretical description of arbitrary individual quantum measurement
processes \cite{GisinPercival1,GisinPercival2,GisinPercival3}.
%change 2 end
It has already
been demonstrated that the QSD method, considered as a dynamical
theory for {\em single quantum systems}, is a valuable tool in the
understanding of the emergence of classical chaos in open quantum
systems~\cite{SpillerRalph94}.  Thus it is expected that it will also
be useful in obtaining new insights into the connection between
classical and quantal behaviour of the driven damped anharmonic
oscillator coupled to a reservoir, a physical system whose classical
dynamics does not exhibit chaos.  This is of particular interest here
as this quantum system does not exhibit hysteresis whereas the
corresponding classical system does.  It will be shown that QSD
provides a mechanics for bistable motion in phase space which is
consistent with the quantum mechanical steady state result that
bistability does not appear for mean values over an
ensemble. Furthermore, QSD helps to understand the crucial role played
by the physical time scales which characterize the approach to the
classical equilibrium positions in phase space and the jumps between
these two classically stable equilibrium positions due to quantum
fluctuations.  In addition, the question is addressed as to what can
be measured, in principle, in an experiment~\cite{TsengGabrielse95} on
such a system and under what conditions such an experimental
observation will yield bistability and hysteresis.  Finally the
assertions of Drummond-Walls and Bortman-Ron are questioned in the
light of QSD.

The paper is organized as follows: In section \ref{QSD}, the QSD model
is briefly depicted. In section \ref{AO} the anharmonic oscillator
model is presented, and a brief review of the classical and quantum
results related to bistability in the steady state is
provided. Section \ref{outlook} describes the anharmonic oscillator
from the point of view of QSD.  Finally, section \ref{Conclu} presents
a discussion of the results and our conclusions.

\section{Quantum state diffusion}\label{QSD}
%===========================================
Open quantum systems are represented by the density operator $\rho$
which evolves in time according to a master equation.  The most
general master equation in the Markov approximation, which preserves
trace and positivity of the density operator $\rho$ can be written in
the Lindblad form~\cite{Lindblad76}
\begin{equation}
\dot{\rho} = -\frac{i}{\hbar}[H,\rho]+\sum_j\left(L_j\rho L_j^\dagger
- \half L_j^\dagger L_j\rho - \half\rho L_j^\dagger L_j \right)
\end{equation}
where $H$ is the Hamiltonian and $L_j$ are Lindblad operators which
represent the effects of the environment on the system.

The quantum state diffusion model (QSD) represents one of the several
possible unravellings of the master equation. According to the QSD
model~\cite{GisinPercival1,GisinPercival2,GisinPercival3,Steimle95,Percival94},
open quantum systems are represented by pure states $|\psi\>$, which
describe individual systems. Evolution of the state $|\psi\>$ is given
by a Langevin-It\^o differential equation
\begin{eqnarray}
|d\psi\> &=& -\frac{i}{\hbar} H|\psi\> dt \nonumber \\ && -
           \half\sum_j (L_j^\dagger L_j +\<L_j^\dagger\>\< L_j\> -
           2\<L_j^\dagger\> L_j )|\psi\> dt \nonumber \\ && + \sum_j
           (L_j -\< L_j\> )|\psi\> d\xi_j .
\end{eqnarray}
The $d\xi_j$ are random differential variables representing
independent complex Wiener processes. They satisfy the following mean
relationships
\begin{equation}
M(d\xi_j)=M(d\xi_j d\xi_k) = 0, \hspace{5mm} M(d\xi_j
d\xi_k^\ast)=\delta_{jk}dt.
\end{equation} 
$M$ represents a mean over an ensemble and $\<
L_j\>=\<\psi|L_j|\psi\>$ the quantum expectation of the operator $L_j$
in the pure state $|\psi\>$.

The QSD trajectories are compatible with the master equation in the
sense that the ensemble average of the projector $|\psi\>\<\psi|$
reproduces the density operator $\rho$:
\begin{equation}
\rho = M(|\psi\>\<\psi |).
\end{equation}
Thus, expectations values $\< A\>_\rho = \mbox{Tr}(\rho A)$ of an
operator $A$ can be computed as the ensemble mean of the quantum
expectations values $\< A\>_\psi$ of the pure state $|\psi\>$.

\section{Anharmonic oscillator}\label{AO}
%==============================

We consider a driven anharmonic oscillator coupled to a thermal
bath~\cite{DW80,HRSW86,Alicki89,MilburnHolmes86}, the temperature of
which is set to zero $(T=0)$ as the prototype model showing
bistability in the classical domain. The damping of this oscillator
with rate $\kappa$, is described by the Lindblad operator
$L=\sqrt{\kappa}a$. The Hamiltonian in a frame rotating with the
frequency $\omega$ of the driving field reads
\begin{equation}
H = \hbar \Domega a^\dagger a +\hbar \beta (a^\dagger + a) +\hbar \chi
(a^\dagger a)^2
\end{equation}
where $a=(\frac{m\omega_0}{2\hbar})^{1/2}
Q+i(\frac{1}{2m\hbar\omega_0})^{1/2}P$ and $a^\dagger$ are the
annihilation and creation operators related to the position $Q$ and
momentum $P$ of the oscillator (m is the mass of the particle).  Here the
parameter $\Domega = \omega_0 - \omega$ measures the detuning between
the eigenfrequency of the oscillator and the driving force. The
parameters $\beta$ and $\chi$ characterize the amplitude of the
driving force the strength of the anharmonicity.  In the following
only positive values of $\chi$ will be considered.

This Hamiltonian whose corresponding classical dynamics is integrable
is well known. It has been used to describe various physical phenomena
such as dispersive optical bistability~\cite{DW80,VogelRisken90},
driven tunneling~\cite{SavageCheng89} and hysteresis in an atomic
system~\cite{BortmanRon94,BortmanRon95} within the framework of the
rotating wave approximation.  
%change3 begin
In the context of optical bistability, for example, the QSD equation
with Hamiltonian (5) and $L=\sqrt{\kappa}a$ describes the continuous
monitoring of the electromagnetic field of frequency $\omega_0$
by individual photon detection processes which involve heterodyning
with a classical intense photon source.
Without the adiabatic approximation 
(i.e. for a time dependent driving $\beta$)
this physical system has been discussed as a model which exhibits 
quantum chaos~\cite{SpillerRalph94}.
%change3 end
It has also
been used to describe more fundamental aspects like the effect of non
linearities in master equations~\cite{HRSW86}, non-Markovian
approximations~\cite{Alicki89}, and more recently in a study of
localization processes~\cite{GarrawayKnight96}.  An appealing feature
of this system is that exact quantum results exist for the
correlations function~\cite{DW80}, the spectrum, and even the
dynamics~\cite{MilburnHolmes86,KartnerSchenzle93} in the absence of
driving.

The classical equivalent for this system exhibits hysteresis in the
steady state~\cite{LL,Kaplan82} while the quantum system does
not. 
%Change 3a 
%Thus, there is a clear discrepancy between the two descriptions.
In the next two sections well known results regarding the classical
and quantum systems are presented in order to make this presentation
self consistent.

\subsection{The classical limit}
The classical equation of motion can be obtained~\cite{DW80} by
factorizing the quantum correlation functions $\< a^\dagger a^2\>
\rightarrow \< a^\dagger\>\< a\>^2$
\begin{equation}
\frac{d\alpha}{dt}= -i\left\{ \beta + (\Domega + \chi)\alpha
+2\chi\alpha^2\alpha^\ast\right\} -\half\kappa\alpha
\label{eqClasicAlpha}
\end{equation}
where $\alpha$ is the mean field amplitude $\alpha = \< a \>$ in the
(semi-) classical limit.

In the steady state regime, the excitation number $|\alpha |^2$ can be
obtained by solving the following equation
\begin{equation}
 |\alpha |^2 = \frac{\beta^2}{(\kappa/2)^2 + \left( \Domega
+\chi+2\chi |\alpha |^2\right)^2}.
\label{EqClassicNorm}
\end{equation}
Once the excitation number $|\alpha |^2$ is known the real and complex
part of the mean amplitude $\alpha$ can be derived using
\begin{equation}
\frac{Re(\alpha)}{|\alpha |^2} = \frac{\Domega+\chi+2\chi |\alpha
|^2}{\beta},
\hspace{1cm} \frac{Im(\alpha)}{|\alpha |^2} = -\frac{\kappa}{2\beta}.
\end{equation}
Expression (\ref{EqClassicNorm}) shows that the classical anharmonic
oscillator can display one, two or three solutions depending on the
choice of parameter values (see figure~\ref{FigHysteresDetuning}).

Provided the following conditions are fulfilled
\begin{eqnarray}
\chi (\Domega + \chi ) &<& 0 \nonumber \\ \left|\frac{\Domega +
\chi}{\kappa/2}\right| &>& \sqrt{3} \label{EqCondHyst} \\ \left[
\frac{27\chi\beta^2}{(\Domega+\chi)^3} +1 +\left(
\frac{3\kappa/2}{\Domega +\chi}\right)^2 \right]^2 &<& \left[ 1-
3\left( \frac{\kappa/2}{\Domega+\chi}\right)^2\right]^3 \nonumber
\end{eqnarray}
then equation (\ref{EqClassicNorm}) has three solutions. Outside of
this range, only one solution is expected.

The first condition expresses the fact that the detuning has to be
oriented in the right direction in order to combine its effect with
that of the anharmonicity. The second shows that detuning and
anharmonicity must be large enough in order to compensate dissipation,
and the third condition gives limiting values for the driving
strength.

Using the theory of linear stability~\cite{DW80,LL}, it can be
verified that when three solutions are present, one of them is always
unstable. Thus the domain of parameters is divided into two regions,
one showing bistability and the other purely mono-stable behaviour
(see figure~\ref{FigHysteresDomain}).  In short, the classical model
exhibits a bistable steady state when the parameters satisfy the above
conditions leading to hysteresis.

\subsection{The quantum limit}
Using the complex $P$ representation, Drummond and Walls~\cite{DW80}
have solved the master equation for the density operator. They
obtained an analytical expression for the moments $\<a^{\dagger
n}a^m\>_\rho$ in the steady state. Notice that in this section, the
brackets $\<\cdots\>_\rho$ represent the ensemble mean of the quantum
mechanical expectation values.  Their result reads
\begin{eqnarray}
\<a^{\dagger n}a^m\>_\rho &=& \left(\frac{z}{2}\right)^{\frac{n+m}{2}}
\frac{\Gamma(c)\Gamma(c^\ast)}{\Gamma(c+m)\Gamma(c^\ast +n)} \nonumber
\\ && \times\frac{F(c+m,c^\ast +n,z)}{F(c,c^\ast,z)}
\end{eqnarray}
where $\Gamma$ is the Gamma function and $F\equiv\ _0\!F_2$ the
generalized Gauss hypergeometric series~\cite{Luke}:
\begin{equation}
_0\!F_2 (c,d,z) = \sum_{n=0}^\infty
\frac{z^n}{n!}\frac{\Gamma(c)\Gamma(d)}{\Gamma(c+n)\Gamma(d+n)}.
\end{equation}
The coefficients $c$ and $z$ depend on the physical parameters in the
following way $c=(\Domega+\chi)/\chi-i\kappa/(2\chi)$ and
$z=2(\beta/\chi)^2$.

The mean excitation number is of particular interest here, it is given
by
\begin{equation}
\< a^\dagger a\>_\rho = \frac{\beta^2}{(\Domega +\chi)^2 +
(\kappa/2)^2}\frac{F(c+1,c^\ast+1,z)}{F(c,c^\ast,z)}.
\label{EqPhotonNumber}
\end{equation}
Using the properties of the hypergeometric series $_0\!F_2$ one can
show that the quantum result is never bistable and thus does not show
any hysteresis (figure~\ref{FigHysteresDetuning}).
 
\section{QSD for the anharmonic oscillator}\label{outlook}
In this section, the problem is tackled using the quantum state
diffusion (QSD) model.  According to QSD, the equation of motion for
the mean field amplitude $\< a\>$ is
\begin{eqnarray}
d\< a\> &=& -i\left[ (\Domega +\chi)\< a\> +\beta +2\chi\< a^\dagger
a^2\>\right] dt -\frac{\kappa}{2}\< a\> dt \nonumber \\ & &
+\sqrt{\kappa}(\< a^2\> - \< a\>^2)d\xi \nonumber \\ & &
+\sqrt{\kappa}(\< a^\dagger a\> -\< a^\dagger\>\< a\>)d\xi^\ast .
\label{EqQSDEvol}
\end{eqnarray}
In this equation, the expectation values are taken in the pure state
$|\psi\>$ describing the evolution along a quantum trajectory.  If one
wants to describe the evolution of the mean value (over an ensemble),
one can take the mean on both sides of the equation (\ref{EqQSDEvol})
to obtain
\begin{equation}
\frac{d\< a\>_\rho}{dt} = -i\left[ (\Domega +\chi)\< a\>_\rho +\beta
+2\chi\< a^\dagger a^2\>_\rho\right] -\frac{\kappa}{2}\< a\>_\rho
\end{equation}
One can check easily that factorizing the quantum correlations in both
of the two preceding equations leads to the classical equation of
motion (\ref{eqClasicAlpha}). Thus, neglecting the quantum
correlations corresponds to ignoring overlapping effects of the wave
packet and quantum fluctuations.

\subsection{Simulation of the dissipative dynamics}
\label{DissipDyn}
The parameters are chosen in the bistable domain. The dynamical
evolution of the QSD equation (\ref{EqQSDEvol}) is computed
numerically using the moving basis or mixed representation simulation
method (MQSD)~\cite{SchackBrunPercival95}. The evolution is computed
over a period of time of the order of $1/\kappa$, {\em the dissipative
time}. In this situation, the system evolves toward a different
``stationary'' state depending on the chosen initial state. The QSD
evolution shows two different limit points where all the trajectories
tend to go after some transient dissipative time. These two points are
called equilibrium points.

When the trajectory starts far away from the two equilibrium points,
it approaches one of them, depending in which basin of attraction it
starts in, rotating with a frequency given by the detuning $\Domega$.
If the wave packet is initially spread out, it tends to
localize~\cite{Steimle95} to a coherent state during the dissipative
transient.  Here the state is said to be localized if its spread is
much smaller than the distance between the two equilibrium
points. Thus the quantum fluctuations have less and less effect.

After the transient damped motion towards one of the equilibrium
points, the wave packet remains localized. This strong localization
property allows one to describe the quantum system in a quasi
classical way.

As a consequence of the anharmonicity, the system does not however
preserve the coherent states to which it is driven by the dissipative
terms. Hence, the quantum correlations never vanish, and the quantum
fluctuations act on the wave packet whose center will fluctuate around
it's equilibrium point.  The dynamics are now dominated by quantum
fluctuations.

These fluctuations have a mean frequency and an amplitude which
depends on the position of the equilibrium point in phase space.  The
fluctuations are bigger for the equilibrium point situated further
away from the origin. This is an effect due to dissipation,
represented by $L=\sqrt{\kappa}a$, which produces a dynamical
behaviour like $\< \dot{a}\> \simeq -\kappa \< a\>$ clearly attracting
the system towards the origin $\< a\>=0$ and not to some local minima.

At this point the QSD trajectories can be roughly seen as classical
trajectories subjected to noise.  There are two distinct equilibrium
points leading to bistable behaviour similar to the classical one.

\subsection{Recovery of the quantum result} 
We know that the QSD model reproduces the quantum result in mean over
an ensemble but if QSD shows bistability, how can the quantum result
be recovered ?  The answer is that the evolution described above is
stable over a very long time compared to the dissipative time, but if
one integrates over a longer time, one sees that any trajectory goes
from the neighborhood of one equilibrium point to the other.  This
transition happens in mean after a time called the {\em transition
time} or {\em exit time}, which can be much longer than the
dissipative time.

In order to observe the transition between the two equilibrium points,
the parameters are chosen such that the maximal excitation number is
set at an intermediate value between the classical and quantum limits.
Also, the integration is now carried over a long time, typically
$10^2$ to $10^4$ times the dissipative time.  The system is initially
set in a coherent state centered far away from the two fixed points
and evolved in time with QSD.  Figure~\ref{FigQSDEvolPS} represents a
trajectory in phase space which shows the dissipative part of the
trajectory, followed by a long period of fluctuations around the
attracting equilibrium point.  This first part of the time evolution
of the trajectory, i.e. its approach to equilibrium, has been
described in the previous section~\ref{DissipDyn}. If the integration
is carried on, the trajectory will suddenly jump to the neighbourhood
of the other equilibrium point.  (The ``jump'' described here is a
diffusive process which allows the quantum trajectory to go from the
basin of attraction of one equilibrium point to the other.)  The
system will remain around the second equilibrium point during some
time and then come back to the first point.  The time spent around
each equilibrium point is such that the quantum expectation value $\<
a\>_\rho$ is recovered in mean.  Due to quantum fluctuations induced
by the coupling to the reservoir the equilibrium points become
metastable.

This transition, which occurs in a time shorter than the dissipative
time, might be viewed as a tunneling process. Savage and
Cheng~\cite{SavageCheng89} have investigated whether bistability can
be associated with quantum superpositions of states in either
well. They have introduced a distinction between coherent and
diffusive mechanisms for quantum tunneling. In our simulations, the
wave packet initially localized in one well becomes delocalized when
it crosses the barrier, making the distinction between these two
mechanisms of tunneling artificial (see
figure~\ref{FigQSDEvolJumpTime}).  Once the barrier is crossed, the
wave packet localizes again.

To confirm the previous description, a mean over an ensemble is
considered. Figure~\ref{FigQSDEvolMeanTime} represents the time
evolution of the mean position and mean momentum. The mean is computed
over 100 trajectories.  The system is initially placed in a coherent
state centered at the classical equilibrium point. This point is
unstable with respect to the other equilibrium point.  The mean
position and momentum evolve, roughly as an exponential decay, to the
quantum stationary values given by the exact quantum result.  For this
typical example, the quantum result is very close to one of the
equilibrium points, because the transition time from the initial to
the final point is much shorter than in the opposite direction.

Hence, the mean result confirms that the initial equilibrium point is
unstable compared to the other one as a consequence of the quantum
fluctuations. This relative instability explains why the quantum
description does not show bistability.  According to QSD even in the
quantum regime, the system is bistable, but the bistability is hidden
by the fluctuations which make the wave packet move from a fairly
localized state in one well to a localized state in the other well.

\section{Discussion}\label{Conclu}
%=================================

We have used QSD to describe the driven damped anharmonic oscillator
in an intermediate regime between quantum and classical.  It has been
shown that states localize along a quantum trajectory and a transition
between the two equilibrium points of the system has been observed.
The localization gives a quantitative justification for the classical
analogy in which a localized particle moves in a double well
potential. This analogy has often been used on a purely qualitative
level without any further justification.  The transition between the
two equilibrium points allows one to recover the quantum result and
reconciles quantum and classical descriptions.  It is worth
emphasizing the following aspects:

\subsection{Transition time}
The QSD model, by introducing quantum fluctuations in an explicit way,
shows explicitly how classical bistability disappears.  Furthermore it
introduces a new time scale, the transition time characteristic of the
transition between the two (classical) equilibrium points.  More
precisely there are two transitions times, the transition time from
one equilibrium point to the other and a different time associated
with the reverse transition. Because one of these transition times is
in general much smaller than the other one, the transition considered
here starts from the less stable equilibrium point which is located
further away from the origin in phase space.

In a sequence of papers, Vogel and Risken~\cite{VogelRisken90} have
calculated the transition rates by solving the equations of motion for
quasidistribution functions using the matrix continued-fraction
method.  They also obtained analytical results for the transition rate
in the limit of large excitation numbers and low damping.

The rate of decay shown on the figure~\ref{FigQSDEvolMeanTime} is an
approximation to the mean transition rate between the two equilibrium
points. The time needed for the decay is clearly much larger than the
dissipative time.

Thus QSD not only gives a qualitative description but can also be used
easily to obtain numerical estimates of the relevant transition
times. We will not address any further the question of the
determination of the transition time in this paper as one can use the
accurate results of Vogel and Risken which confirm the possibility of
very large transition times compared to the dissipative time.

\subsection{Ideal experiment}
Let us consider the following ideal experiment :
%\change4a begin 
(see \cite{TsengGabrielse95} for a practical realization)
%\change4a end 
In order to see
hysteresis a single quantum mechanical anharmonic oscillator is
measured continuously under conditions in which its classical
counterpart would be bistable.  Let us assume that the driving
frequency $\omega$ is varied step by step from low to high frequencies
and reversed, spanning twice the classically bistable domain.  Once
the frequency is modified, the experimenter waits a time, called {\em
the measurement delay} $t_m$, which is assumed to be much longer than
the dissipative time, before doing any further measurement.  The
excitation number of the oscillator is measured before changing the
frequency of the driving force again.  Thus this type of resonance
experiment corresponds to an adiabatic sweeping of the frequency with
respect to the ``fast'' dissipative dynamics.  Furthermore, let us
assume that this continuous measurement is ideal in the sense that the
excitation number of the harmonic oscillator can be measured
nondestructively and that it does not perturb the systems transition
rates, i.e. it is a nondemolition measurement.
%change4 begin 
Such a measurement can be performed, for example, in an optically
bistable system by monitoring the state of the field mode by
heterodyning with an intense classical photon source. Alternatively,
this measurement can also be realized by observing the relativistic
motion of a
resonantly driven electron in a Penning trap, like in the
recently performed experiment of Ref. \cite{TsengGabrielse95}.
%change4 end 

According to QSD such a measurement should reproduce the curve
depicted in figure~\ref{FigQSDIdealExperiment} showing hysteresis.
The experimenter should obtain such a curve fluctuating around one of
the two classical steady state values for a while and then jumping to
the other value. The combination of the two jumps occurring when the
driving frequency is ramped from low to high frequencies and reversed
allows one to define the detuning width $\Delta\Omega$ as the size of
the bistable region.  Figure~\ref{FigQSDIdealExperiment} represents
such a result and shows the detuning width $\Delta\Omega$ for this
particular realization.  The detuning width is different for each
realization of this experiment, the transition being a stochastic
event.
 
An experiment carried out in the classical limit does not show any
fluctuations and the two jumps occur always at the same detuning
value.  In this case the detuning width $\Delta\Omega$ corresponds to
the full size of the bistable region.

If one uses the density matrix to describe such an ideal experiment,
the result will also show two distinct transitions.  The mean detuning
width depends not only on the characteristic physical parameters of
the driven damped anharmonic oscillator but also on the measurement
delay $t_m$.  If the mean transition time $\tau$ is much larger than
the dissipative time, i.e.  $ \tau \gg 1/\kappa$, one can distinguish
between the two limiting cases: (1) If the measurement delay is small
relative to the transition time, i.e.  $ \tau \gg t_m$, then the mean
detuning width $\Delta\Omega$ has a finite value, showing
bistability. (2) At the opposite extreme, i.e. for $ t_m \gg \tau$,
the detuning width is equal to zero, showing no hysteresis at all, in
agreement with the quantum steady state result.

\subsection{Bistability and the classical limit}
The classical limit is valid for high excitation number.  In this
limit the mean transition time is so large and the transition between
the two equilibrium points so infrequent that they can be
neglected. In this limit any finite observational time satisfies the
conditions for the experimental observation of hysteresis.

When the excitation number decreases sufficiently for the mean
transition time to take a very large but accessible value, one has to
distinguish between the cases where the measurement delay is smaller
or larger than the transition time.  The classical description is
still valid in the former case but does not apply anymore in the
latter.  The fluctuations have to be taken into account for a correct
description of this situation.

If we continue lowering the excitation number, still keeping the mean
transition time large compared to the dissipative relaxation time,
then the classical theory no longer gives a good description of the
dynamics since even for a measurement delay much smaller than the
transition time it predicts a fixed detuning width.  If one uses
quantum theory, which includes the fluctuations, one will be able to
obtain the correct behaviour.

Finally when the excitation number is small, the classical theory is
no longer valid. One has to use the quantum theory and specify the
measurement delay in order to describe correctly the result of an
experiment. The bistability is not destroyed in any of these cases but
it is simply hidden by the quantum fluctuations.

This situation is very similar to that of a classical driven
anharmonic oscillator coupled with a thermal bath with non-zero
temperature. Introducing thermal fluctuations also hides the
bistability of the steady state and introduces a classical transition
time (see \cite{DykmanKrivoglaz80}). In order to observe bistability,
one has to introduce a measurement delay much larger than the
relaxation time in the absence of thermal fluctuations, but shorter
than the mean transition time.  The thermal fluctuation can be
neglected only when the transitions take place in a time much larger
than the observational time.

In all the previous situations the quantum theory applies.  Because
the density matrix automatically includes the mean over an ensemble,
there is no clear distinction between the dissipative dynamics and the
dynamics induced by the fluctuations. QSD, by unraveling the different
quantum trajectories, helps one to understand the role played by the
statistical mean.

\section{Summary}

We have shown that QSD leads to quantum trajectories which exhibit
bistability for the driven damped anharmonic oscillator and that the
quantum steady state result is recovered through random switching
between the two equilibrium points due to quantum fluctuations.  The
fluctuations also introduce a characteristic time scale: the mean
transition time.

We have also shown that it is still possible to observe bistability
dynamically in this quantum system by introducing a measurement delay
$t_m$.  An experiment will show hysteresis only if the transition time
between the (classical) equilibrium points, $\tau$, is much larger
than all other relaxation times involved (approximately $1/\kappa$).
If the transition time is not that large, it is not possible to
observe any hysteresis effects and the quantum steady state solution
is expected.  Furthermore, provided the transition time is much larger
than the other times, hysteresis can be seen only if a measurement
delay is such that $t_m \ll \tau$.  Thus, within this interpretation
the classical result is valid only if the quantum fluctuations are so
small that they induce a transition time very large compared to the
period of observation.

Our results confirm the statistical description given by Drummond and
Walls.  The results of Bortman and Ron have to be examined more
carefully.  Strictly speaking, even for low excitation number, the
bistability is not destroyed, it is just hidden by the quantum
fluctuations. But one cannot neglect the effect of the fluctuations
upon the two stable states since it is exactly these fluctuations
which prevent an experimental observation of hysteresis in the steady
state.

Finally, if QSD is used to describe the dynamical behaviour of a
single quantum system our investigations might be of particular
relevance for 
%change5 begin
experiments on optical bistability or for
%change5 end
the recently performed experiments of Gabrielse et al.
\cite{TsengGabrielse95} in which hysteresis was observed.

\section*{Acknowledgments}
%========================
It is a pleasure to acknowledge very stimulating discussions with
Todd Brun, Nicolas Gisin and R\"udiger Schack. Financial support by
the EU under its Human Capital and Mobility Programme and the Deutsche
Forschungsgemeinschaft is gratefully acknowledged. FMF and PFOM also
acknowledge financial support from the PRAXIS XXI programme of the
Junta Nacional de Investigacao Cientifica e Tecnologica (JNICT), Portugal.

\newpage
% List of Figures
%================

\begin{figure}
\caption{Steady state excitation number $|\alpha|^2$ 
(classical result - dots) and $\< a^\dagger a\>$ 
(quantum result - line) versus the detuning $\Domega$. The 
parameters are: the damping $\kappa=1.5$, the driving $\beta=-7.0$ 
(choosing  $\beta$ real is equivalent to setting the origin for the phase 
of $\< a\>$, introducing a complex $\beta$ simply produces a rotation in 
phase space)
and the anharmonicity $\chi=0.05$.}
\label{FigHysteresDetuning}
\end{figure}

\begin{figure}
\caption{Representation of the domain of hysteresis.
The three conditions given by Eqn.(\ref{EqCondHyst})  
in the text describe the border of this domain.
Inside the bounded region the system possesses three solutions, two stable 
and one unstable, this is the bistable domain. Outside, the system has 
only one solution, always stable. Notice that the bistable region is 
entirely defined by only two parameters
$x=\frac{\kappa/2}{\Domega+\chi}$ 
and
$y=\frac{(\kappa/2)^3}{\beta^2\chi}$.}
\label{FigHysteresDomain}
\end{figure}

\begin{figure}
\caption{Evolution of a quantum trajectory in phase space. The parameters 
are: $\kappa=1.5$, $\beta=-7.0$, $\chi=0.05$ and the detuning $\Domega = 
-5.0$. The initial state is chosen to be a coherent state
 centered at $(\< Q\>=7,\< P\>=14)$. 
The first stage of the evolution is the decay towards a local minimum, in a 
time of the order of $1/\kappa$. Then the system fluctuates around the 
equilibrium point for an amount of time given, in mean, by the transition 
time. 
Finally, a big enough fluctuation occurs to project the system into the 
other basin of 
attraction where the system remains for a very long time.}
\label{FigQSDEvolPS}
\end{figure}

\begin{figure}
\caption{Representation in time of A) the position $\< Q\>$ (full line) 
and its variance $\Delta Q^2$ (dotted line)  and B) the momentum $\< P\>$ 
(full line) and its variance $\Delta P^2$ (dotted line) at the particular 
instant of the transition (approximatively at $t=596.3$ in this example).
Same parameters as figure~\ref{FigQSDEvolPS}.
Notice the delocalization in space of the wave packet at the transition. 
Before and after the transition, the variances $\Delta Q^2$ and $\Delta P^2$
are small (compared to the distance between equilibrium points) showing
localized states.}

\label{FigQSDEvolJumpTime}
\end{figure}

\begin{figure}
\caption{Evolution in time of the mean position $\< Q\>$ (lower curve) 
and momentum $\< P\>$ (upper curve). 
Parameters same as figure~\ref{FigQSDEvolPS}. The mean is taken over 100 
realizations. The time scale of the decay is much larger than the 
dissipative time of $1/\kappa = 0.67$.}
\label{FigQSDEvolMeanTime}
\end{figure}

\begin{figure}
\caption{Simulation of an ideal (single) experiment according to the QSD model. 
Parameters are the same as in figure~\ref{FigHysteresDetuning}. 
The detuning step is 0.1 and the measurement time is $t_m=50$. 
The dotted line represents the classical steady state excitation number.}
\label{FigQSDIdealExperiment}
\end{figure}

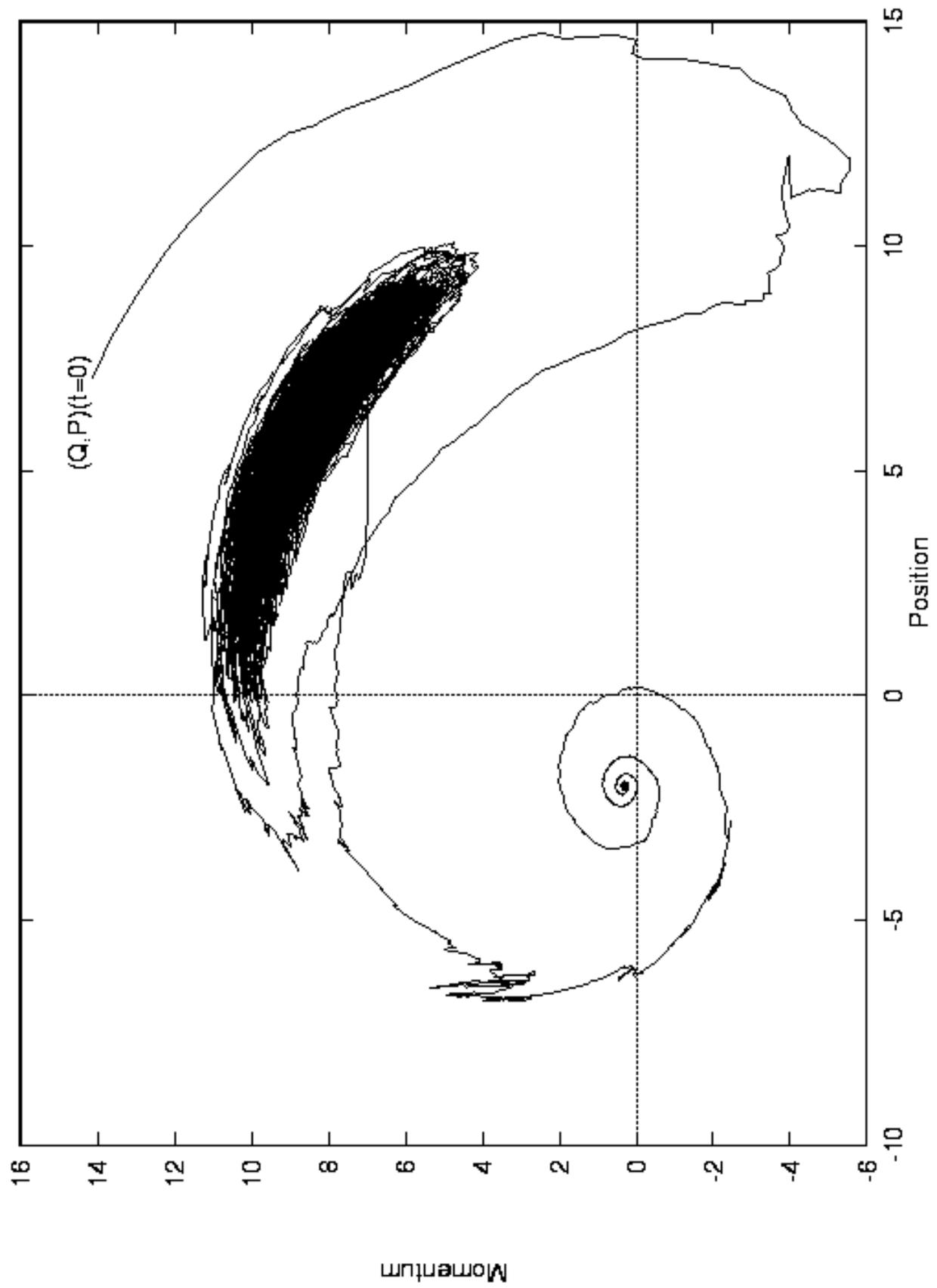
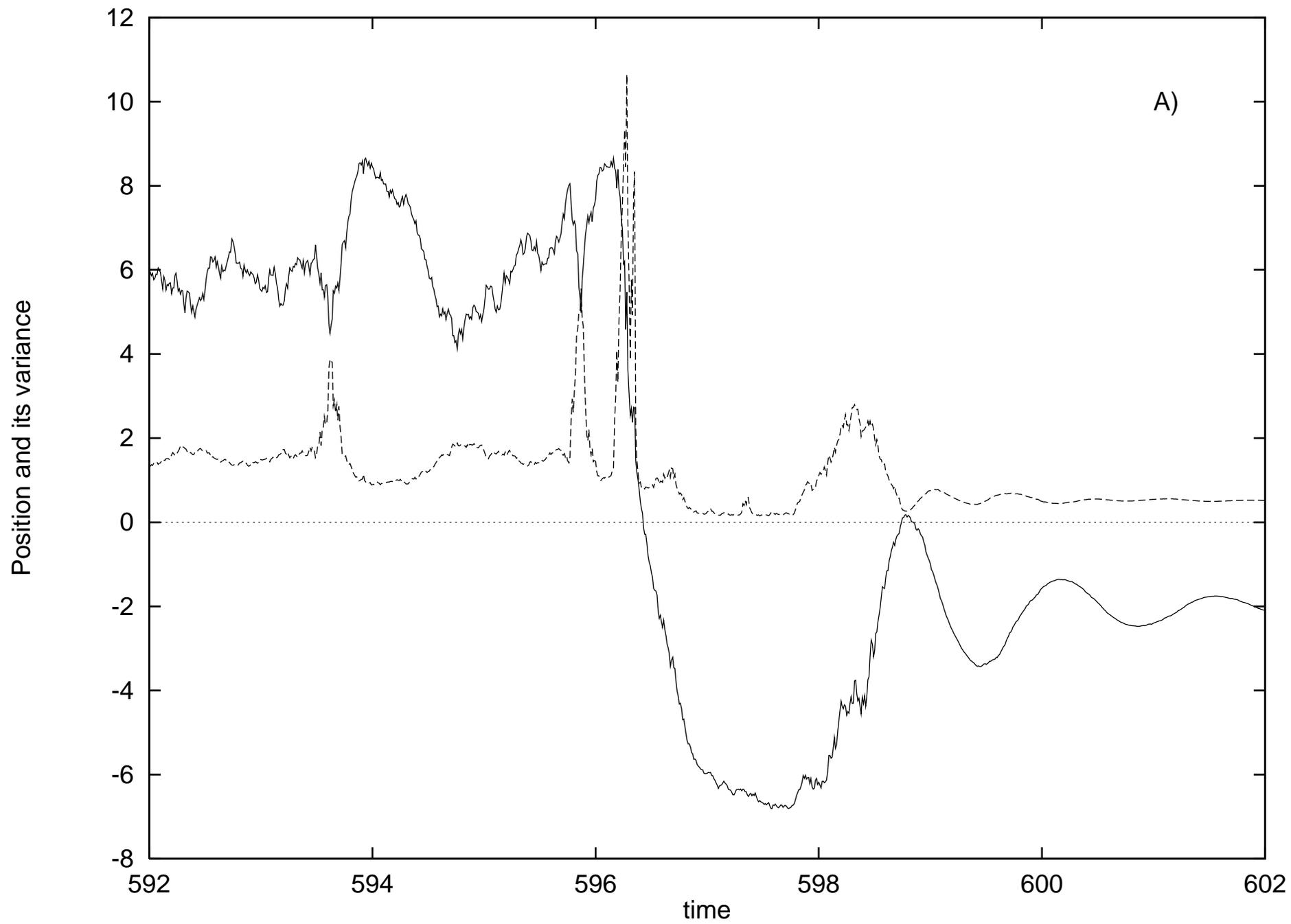
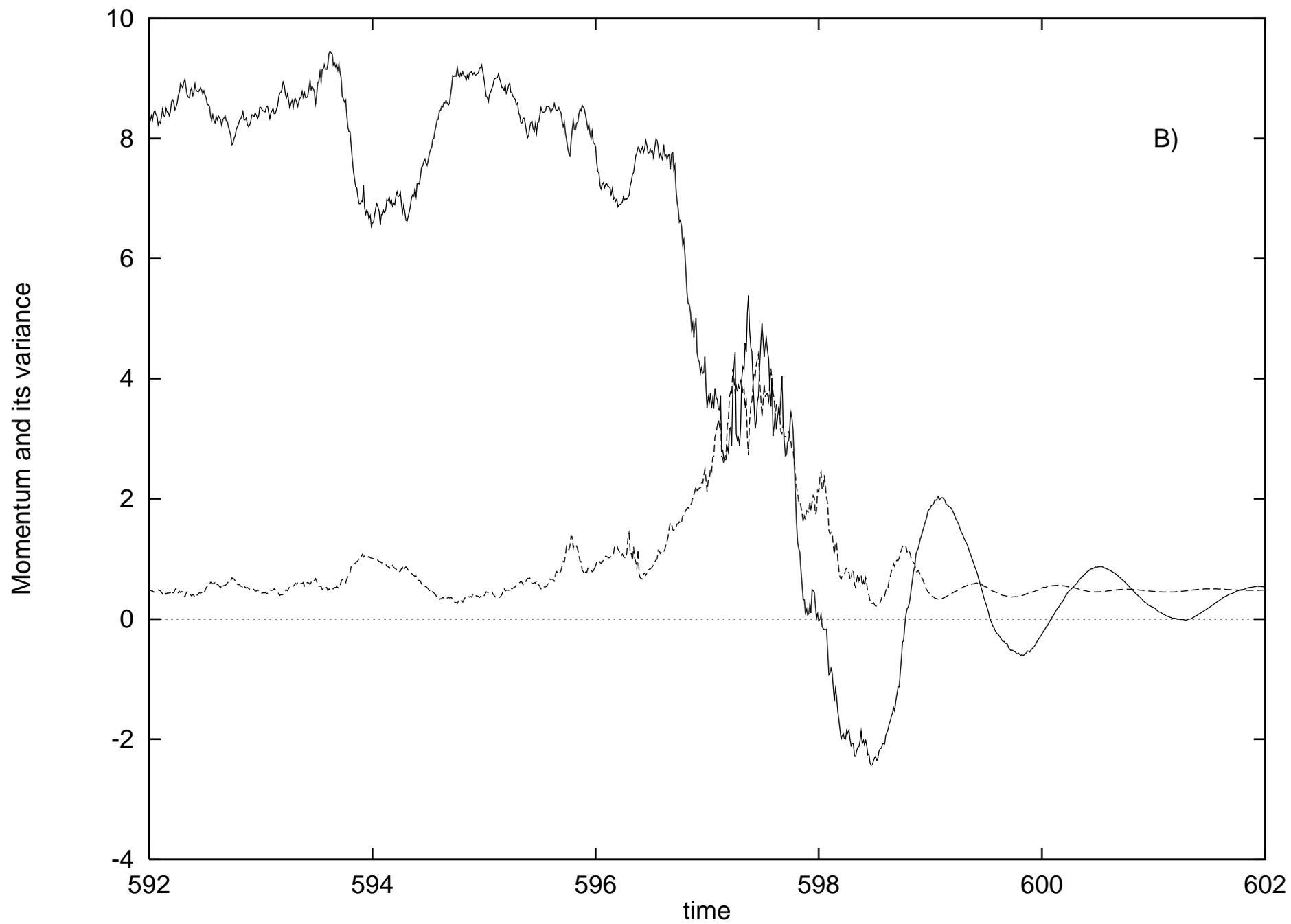
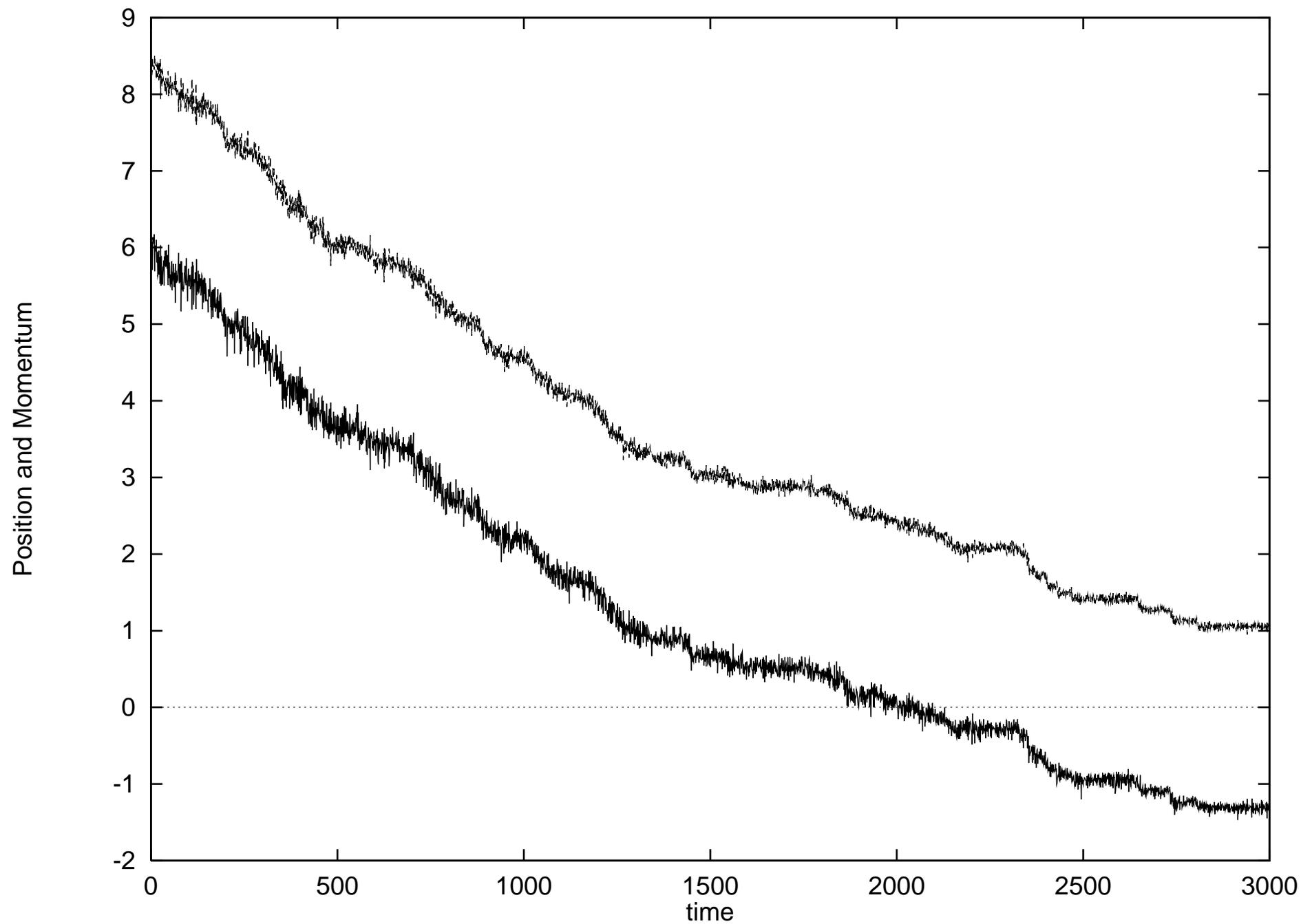
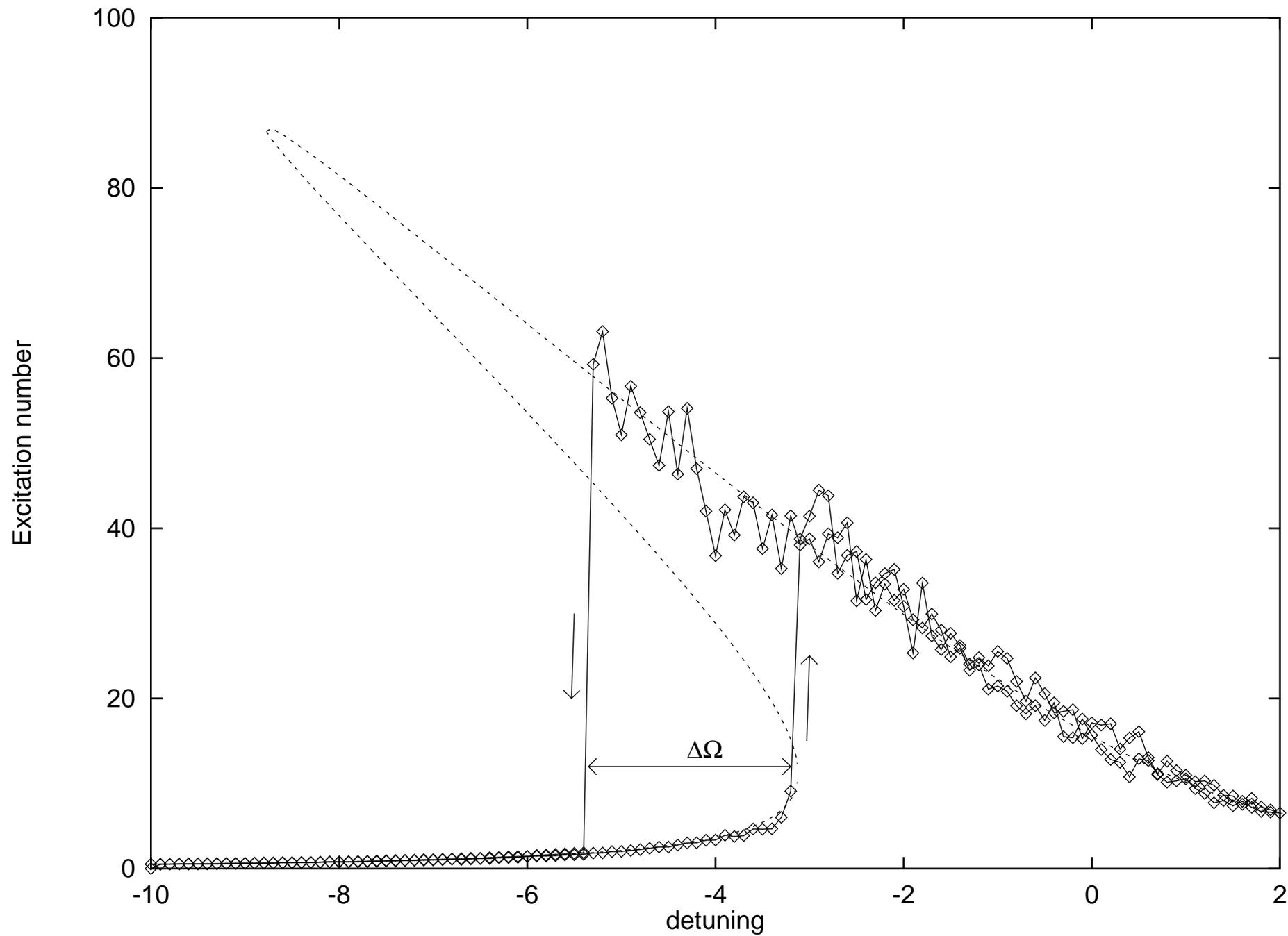
\end{document}